\documentclass[conference, 10pt]{IEEEtran}
\pdfoutput=1
\IEEEoverridecommandlockouts
\usepackage{cite}
\usepackage{epsf}
\usepackage{psfrag}
\usepackage{amsmath,amssymb,amsfonts}
\usepackage{algorithmic}
\usepackage{graphicx}
\usepackage{textcomp}
\usepackage{xcolor}
\newcommand{\overbar}[1]{\mkern 1.5mu\overline{\mkern-1.5mu#1\mkern-1.5mu}\mkern 1.5mu}
\def\BibTeX{{\rm B\kern-.05em{\sc i\kern-.025em b}\kern-.08em
		T\kern-.1667em\lower.7ex\hbox{E}\kern-.125emX}}
\newtheorem{theorem}{Theorem}

\newtheorem{lemma}{Lemma}
\newtheorem{definition}{Definition}
\newtheorem{remark}{Remark}
\begin{document}
	
	\title{Empirical Coordination Subject to a Fidelity Criterion\\
		{\footnotesize \textsuperscript{}}}
	%\begin{figure}
		%\centering
	%	\includegraphics[width=0.7\linewidth]{"Imperfect Empirical Coordination"}
	%	\caption{}
	%	\label{fig:imperfect-empirical-coordination}
	%\end{figure}
	%}
	
	\author{\IEEEauthorblockN{Michail Mylonakis}
		\IEEEauthorblockA{\textit{Division of Inf. Science \& Eng.} \\
			\textit{KTH Royal Institute of Technology}\\
			%Stockholm, Sweden \\
			mmyl@kth.se}
		\and
		\IEEEauthorblockN{ Photios A. Stavrou}
		\IEEEauthorblockA{\textit{Division of Inf. Science \& Eng. } \\
			\textit{KTH Royal Institute of Technology}\\
			%Stockholm, Sweden \\
			fstavrou@kth.se}
		\and
		\IEEEauthorblockN{ Mikael Skoglund}
		\IEEEauthorblockA{\textit{Division of Inf. Science \& Eng.} \\
			\textit{KTH Royal Institute of Technology}\\
			%Stockholm, Sweden \\
			skoglund@kth.se}}
	%\and
	%\and
	%\and
	%\and
	
	\maketitle
	
	\begin{abstract}
	
		We study the problem of empirical coordination subject to a fidelity criterion for a general set-up. We prove a result which indicates a strong connection between our framework and the framework of empirical coordination developed in \cite{cuff:2010}. It turns out that when we design codes that achieve empirical coordination according to a given distribution and subject to the fidelity criterion, it is sufficient to consider codes that produce actions of the same joint type for a class of types which is close enough to our desired distribution is some sense.
		
			%there is not a more efficient way of achieving 
		%Specifically, when we design codes which achieve empirical coordination according to a given distribution subject to our fidelity criterion, it's optimal to restrict our search in the class of codes that achieve empirical coordination (as defined in \cite{cuff:2010}) according to any distribution from a specific class of distributions which are close to our desired distribution in some sense.  See figure \ref{fig:fig1}.
		%Finally, we apply the result in some simple set-ups.
		%establish formally and discuss a general coordination capacity result which 
		%brings close the problems of perfect and imperfect empirical coordination for any set-up.
	\end{abstract}

	\section{Introduction}

\begin{figure}
 	\begin{center}
 		
 		\includegraphics[width=7cm,height=6cm,keepaspectratio]{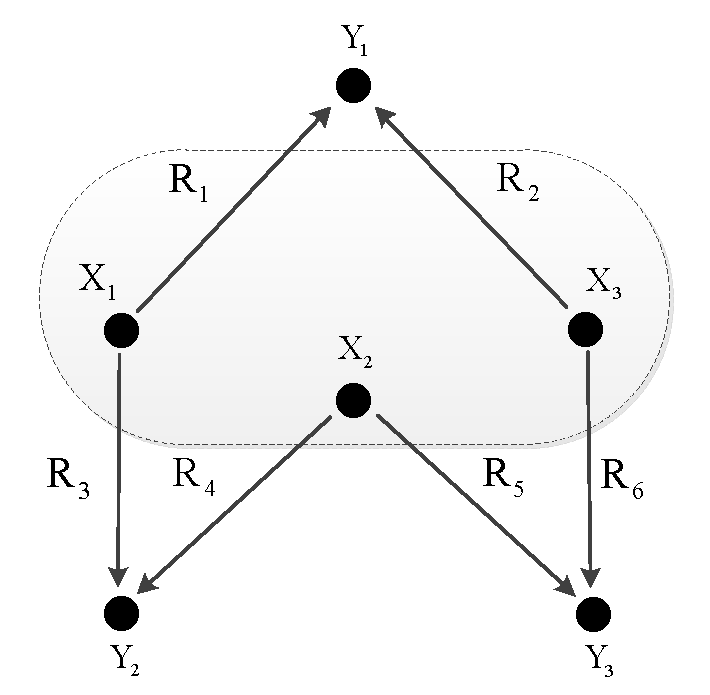}
 	\end{center}
 	\caption{An illustrative example of the considered framework. The nodes in this network have rate-limited communication links between them. 
 		In this example, actions $X_1, X_2$, and $X_3$ are assigned  according to the joint distribution $p_0(x_1, x_2, x_3)$. Then, using the communication that is available to all nodes, the actions $Y_1, Y_2$, and $Y_3$ are produced. We ask, the amount of the communication that is required such that the average distance between the joint type of the actions and the desired distribution to be smaller than a certain level $\Delta$.}
 	%	Each node performs an action where some of these actions are selected randomly by nature. 
 	\label{fig:fig1}
 \end{figure}

Communication is one of the most important and expensive resources in a network with nodes who desire to establish cooperative behavior. When relevant information is known only at some nodes, finding the minimum communication requirements to coordinate actions can be posed as a network information theory problem. Specifically, we consider the communication needed to establish coordination summarized by a joint probability distribution of behavior among all nodes in the network (see Fig. \ref{fig:fig1}).

Cuff {\it et al.} in \cite{cuff:2010} introduced two different notions of coordination, empirical and strong.
According to \cite{cuff:2010}, empirical coordination is established if the joint type of the actions in the network is close to the desired distribution. This kind of coordination has been studied in various set-ups (see e.g., \cite{bereyhi:2013,letreust:2015a,letreust:2015b}) and it has been combined with ideas from other fields like game theory (see e.g.,\cite{letreust:2016}). Strong coordination instead deals with the joint probability distribution of the actions. If the actions in the network are generated randomly so that a statistician cannot reliably distinguish (as measured by total variation) between the constructed $n$-length sequence of actions and random samples from the desired distribution, then strong coordination is achieved. The literature related to strong coordination is vast and includes more complex set-ups, such as extensions in networks with noisy channels, applications in power control etc, (see, e.g., \cite{vellambi:2018,cervia:2017,larrousse:2018}).
It should be noted that \cite{cuff:2010} establishes a fundamental difference between empirical and strong coordination regarding the impact that common randomness can have to the accomplishment of the coordinated behaviour. Specifically, it turns out that common randomness does not play a necessary role in achieving empirical
 coordination but it is a valuable resource for achieving strong coordination. For more details on the foundation of common randomness, see \cite{wyner:1975}.
 
In this paper, we deal with empirical coordination. However, instead of asking the joint type of the actions to be close to the desired distribution in probability, we require something less restrictive. We demand the average distance (as measured by total variation) between the joint type of the actions and the desired distribution to be smaller than a certain level $\Delta$ for long enough $n$-length sequences. To distinguish between the two kinds of empirical coordination, i.e., ours and the one in \cite{cuff:2010}, we will call the one established in \cite{cuff:2010} \emph{perfect} empirical coordination and ours \emph{imperfect} empirical coordination.
This kind of imperfect empirical coordination was first introduced and studied for a specific set-up (including a more general class of fidelity metrics) by Kramer and Savari in \cite{kramer:2007}. 
Here, we generalize this framework (in the case of the total variation fidelity metric) for a general network setting. We establish formally a coordination result which shows that when our task is to design a good coordination code that achieves imperfect empirical coordination according to some distribution, it is optimal to restrict our search exactly to the class of codes that achieve perfect empirical coordination according to any distribution from a class of distributions which is close enough to our desired distribution with respect to the total variation distance.  %$N_{\Delta}\big(p_0\left(x\right)p_{YZ|X}\left(y,z|x\right)\big)$: 
%which shows that codes who achieve perfect empirical coordination to some carefully chosen distributions, are also optimal for the problem of imperfect empirical coordination. Finally, we apply the capacity result in some simple set-ups.

%As a consequence, the problem of designing a good coordination code under an average distortion criterion can be reducted, without loss of optimality, to the problem of designing a coordination code who achieves perfect empirical coordination according to a carefully chosen distribution.

%The paper is organized as follows. In Section II, we give all the required definitions. In Section III, we state, prove and interpret our main result. In Section IV, we apply the main result in some simple set-ups to get the capacity regions for the problem of imperfect empirical coordination. We conclude our work in Section V. 

\section{Definitions}
\begin{figure}
	\begin{center}
		
		\includegraphics[width=7cm,height=4cm,keepaspectratio]{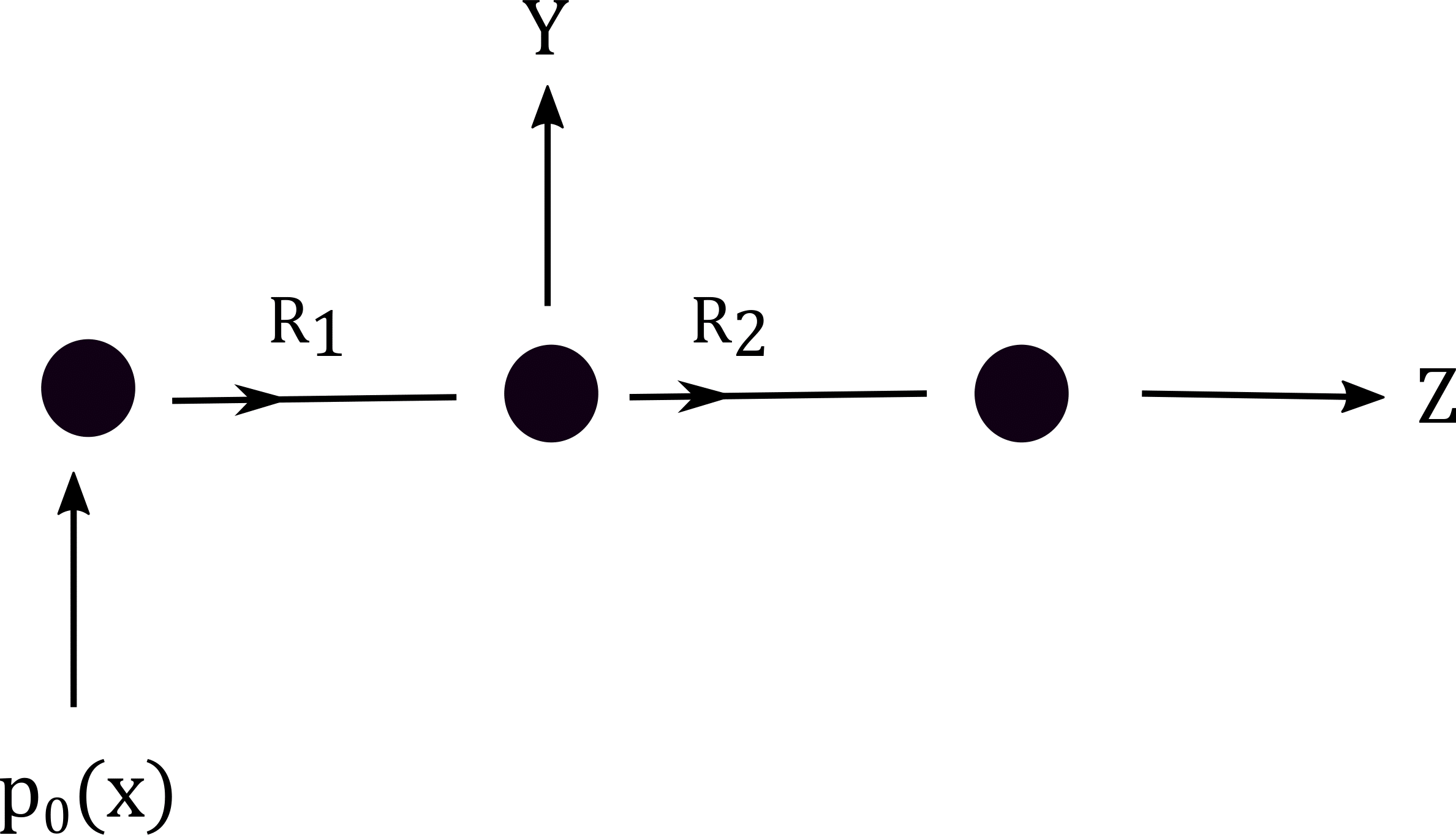}
	\caption{To simplify things, we define formally our problems and we state our general result in the context of the cascade set-up of the figure.}	
	\label{fig:fig2}	
	\end{center}
	
\end{figure}

In this section, we state the definitions of perfect and imperfect empirical coordination in the context of the cascade network of Fig \ref{fig:fig2}. These definitions have obvious generalizations to other networks. We begin with some basic mathematical concepts and the definition of the $\Delta$-neighborhood, a concept which will help us in the statement of our results.
\begin{definition}[Joint type] The joint type $P_{x^n,,y^n,z^n}$ of a tuple of sequences $\left(x^n,y^n,z^n\right)$ is the empirical probability mass function, given by
	\begin{equation*}P_{x^n,y^n,z^n}\left(x,y,z\right)\triangleq \frac{1}{n}\sum_{i=1}^n{\mathbf 1\big(\left(x_i,y_i,z_i\right)=\left(x,y,z\right)\big)},\end{equation*}
	for all $\left(x,y,z\right)\in \mathbb{X}\times \mathbb{Y}\times \mathbb{Z}$, where $\mathbf 1$ is the indicator function.
\end{definition}

\begin{definition}[Total variation]The total variation between two probability mass functions is given by \begin{equation*}\|p\left(x,y,z\right)-q\left(x,y,z\right)\|_{TV}\triangleq\frac{1}{2}\sum_{x,y,z}{|p\left(x,y,z\right)-q\left(x,y,z\right)|}.\end{equation*}
\end{definition}
\begin{definition}[$\Delta$-neighborhood]
	The $\Delta$-neighborhood of a distribution $p\left(x,y,z\right)$ is defined as
	\begin{IEEEeqnarray*}{rCl}\IEEEeqnarraymulticol{3}{l} {N_{\Delta}\big(p\left(x,y,z\right)\big)}\\&\triangleq& \big\{q(x,y,z):\|p\left(x,y,z\right)-q\left(x,y,z\right)\|_{TV}\leq\Delta\big\}.
	\end{IEEEeqnarray*}
\end{definition}
%\subsection{Coordination Code}
A $\left(2^{nR_1},2^{nR_2},n\right)$ coordination code is the protocol which is used  to coordinate the actions in the network for a block
of n time periods. The coordination code and the distribution
of the random actions, $X_n$, induce a joint distribution on
the actions in the network.
\begin{definition}[Coordination code\cite{cuff:2010}]The $\left(2^{nR_1},2^{nR_2},n\right)$ coordination code for our set-up consist of four functions-an encoding function 
	\begin{equation*}
	i:\mathbb X^n%\times\Omega%
	\rightarrow\left\{1,\dots,2^{nR_1}\right\},
	\end{equation*}
	a recoding function
	\begin{equation*}j:\left\{1,\dots,2^{nR_1}\right\}
	\rightarrow\left\{1,\dots,2^{nR_2}\right\},
	\end{equation*}
	and two decoding functions 
	\begin{align*}
	y^n&:\left\{1,\dots,2^{nR_1}\right\}%\times\Omega%
	\rightarrow \mathbb Y^n,\\
	z^n&:\left\{1,\dots,2^{nR_2}\right\}%\times\Omega%
	\rightarrow \mathbb Z^n.
	\end{align*} 
\end{definition}
\begin{definition}[Induced distribution\cite{cuff:2010}]
	The induced distribution $\tilde{p}\left(x^n, y^n, z^n\right)$ is the resulting joint distribution of the actions in the network $X^n, Y^n$, and $Z^n$ when a $\left(2^{nR_1},2^{nR_2},n\right)$ coordination code is used.
\end{definition}

In our set-up, the actions $X^n$ are chosen by nature to be i.i.d according to $p_0\left(x\right)$. Thus, $X^n$ is distributed according to a product distribution \begin{equation*} X^n\sim \prod_{i=1}^{n}{p_0\left(x_i\right)}. \end{equation*}
%Moreover, define $p_{0x}\left(x\right)=\sum_{\hat{x}}{p_0\left(x,\hat{x}\right)}$.
The actions $Y^n$ and $Z^n$ are  functions of $X^n$ given by implementing the coordination code as \begin{align*} Y^n&=y^n\Big(i\left(X^n\right)\Big),\\Z^n&=z^n\Big(j\big(i\left(X^n\right)\big)\Big).\end{align*}	
	%\begin{IEEEkeywords}
	%empirical coordination
	%\end{IEEEkeywords}

%\subsection{Perfect and imperfect empirical coordination}
Perfect empirical coordination is achieved if the joint type of the actions in the network tends to the desired distribution with high probability.
\begin{definition}[Achievability for perfect coordination\cite{cuff:2010}] A desired distribution $p_{X,Y,Z}\left(x,y,z\right)\triangleq p_{0}\left(x\right)p_{YZ|X}\left(y,z|x\right)$ is achievable for empirical coordination with the rate-pair $\left(R_1,R_2\right)$ if there exists a sequence of $\Big(2^{nR_1},2^{nR_2},n\Big)$  coordination codes such that 
	%if for arbitrary $\epsilon_1,\epsilon_2\geq 0$ there exists a sequence of $\Big(e^{n\left(R_1+\epsilon_1\right)},e^{n\left(R_2+\epsilon_2\right)},n\Big)$  coordination codes such that 
	
	%the total variation between the joint type of the actions in the network and the desired distribution goes to zero in probability (under the induced distribution). That is,
	\begin{equation*}\|P_{x^n,y^n,z^n}\left(x,y,z\right)-p_{0}\left(x\right)p_{Y,Z|X}\left(y,z|x\right)\|_{TV}\to 0,
	\end{equation*}
	in probability.	\label{def:perfect}	
\end{definition}
\begin{definition}[Coordination capacity region\cite{cuff:2010}]
	The coordination capacity region $C_{p_0}^P$ for the source distribution $p_0\left(x\right)$ is the closure of the set of rate-coordination tuples $\big(R_1,R_2,p_{Y,Z|X}\left(y,z|x\right)\big)$ that are achievable:
	\begin{equation*}
	C_{p_0}^P\triangleq
	\mathbf {Cl}\left\{ \,
	\begin{IEEEeqnarraybox}[
	\IEEEeqnarraystrutmode
	\IEEEeqnarraystrutsizeadd{1pt}
	{1pt}][c]{l}
	\big(R_1,R_2,p_{Y,Z|X}\left(y,z|x\right)\big):\\
	p_{0}\left(x\right)p_{Y,Z|X}\left(y,z|x\right)\\\text{is achievable at rates $\left(R_1,R_2\right)$}
	\end{IEEEeqnarraybox}\right\}.
	\end{equation*}
	%\begin{equation*}C_{p_0}^P\triangleq \mathbf {Cl}\Big\{\big(R_1,R_2,p_{Y,Z|X}\left(y,z|x\right)\big):p_{0}\left(x\right)p_{Y,Z|X}\left(y,z|x\right)\text{is achievable at rates $\left(R_1,R_2\right)$}\Big\}.\end{equation*} 
\end{definition}
%\subsection{Imperfect Empirical Coordination}
Imperfect empirical coordination is achieved if the average distance between the joint type of the actions in the network and the desired distribution is under a certain level $\Delta$. We will call this kind of coordination $\Delta$-empirical coordination.
\begin{definition}[Achievability for  $\Delta$-empirical  coordination] A desired distribution $p_{X,Y,Z}\left(x,y,z\right)\triangleq p_{0}\left(x\right)p_{Y,Z|X}\left(y,z|x\right)$ is achievable for $\Delta$-empirical coordination with the rate-pair $\left(R_1,R_2\right)$ if there is an $N$ such that for all $n>N$, there exists a coordination code $\Big(2^{nR_1},2^{nR_2},n\Big)$ such that
	%if for arbitrary $\epsilon \geq 0$ and large enough $n$, there exists a coordination code $\Big(2^{nR_1},2^{nR_2},n\Big)$ such that
	\begin{equation*}
	%\uplim_{n \to \infty}
	\mathbb{E}\big\{\|P_{x^n,y^n,z^n}\left(x,y,z\right)-p_{0}\left(x\right)p_{YZ|X}\left(y,z|x\right)\|_{TV}\big\}\leq \Delta.\end{equation*}\label{def:imperfect}
\end{definition}
%\begin{definition}
%The coordination capacity region $C_{p_0}^I$ for the source distribution $p_0\left(x\right)$ is the closure of the set of rate-coordination-distortion tuples $\big(R_1,R_2,\Delta,p_{YZ|X}\left(y,z|x\right)\big)$  that are achievable:

%\begin{multline*}
%C_{p_0}^I\triangleq
%\left.
%\mathbf {Cl}\left\{ \,
%\begin{IEEEeqnarraybox}[
%5\IEEEeqnarraystrutmode
%\IEEEeqnarraystrutsizeadd{7pt}
%{7pt}][c]{l}
%\big(R_1,R_2,\Delta,p_{YZ|X}\left(y,z|x\right)\big):\\
%p_{0}\left(x\right)p_{YZ|X}\left(y,z|x\right)\text{is $\Delta$-achievable}\\\text{at rates $\left(R_1,R_2\right)$}
%\end{IEEEeqnarraybox}\right\}.
%\right.
%\end{multline*}

%\begin{equation*}C_{p_0}^I\triangleq \mathbf {Cl}\Big\{\big(R_1,R_2,\Delta,p_{YZ|X}\left(y,z|x\right)\big):p_{0}\left(x\right)p_{YZ|X}\left(y,z|x\right)\text{is $\Delta$-achievable at rates $\left(R_1,R_2\right)$}\Big\}.\end{equation*}
%\end{definition}

\begin{definition}[Rate-distortion-coordination region]
	The rate-distortion-coordination region $R_{p_0}^I$ for the source distribution $p_0\left(x\right)$ and for a fixed conditional distribution $p_{Y,Z|X}\left(y,z|x\right)$ is defined as:
	
	\begin{multline*}
	R_{p_0}^I\big(p_{Y,Z|X}\left(y,z|x\right)\big)\\\quad \triangleq
	\mathbf{Cl}\left.
	\left\{ \,
	\begin{IEEEeqnarraybox}[
	\IEEEeqnarraystrutmode
	\IEEEeqnarraystrutsizeadd{1pt}
	{1pt}][c]{l}
	\left(R_1,R_2,\Delta\right):\\
	p_{0}\left(x\right)p_{Y,Z|X}\left(y,z|x\right) \text{is achievable}\\\text{for $\Delta$-empirical coordination at rates $\left(R_1,R_2\right)$}
	\end{IEEEeqnarraybox}\right\}.
	\right.
	\end{multline*}
	%\begin{equation*}R_{p_0}^I\big(p_{YZ|X}\left(y,z|x\right)\big)\triangleq \Big\{\left(R_1,R_2,\Delta\right):\big(R_1,R_2,\Delta,p_{YZ|X}\left(y,z|x\right)\big)\in C_{p_0}^I\Big\}.\end{equation*}
\end{definition}
\section{Main result}	
In this section, we state and prove the main result. For simplicity, we state and prove the result in the case of the cascade set-up of Fig. \ref{fig:fig2}. Nevertheless, the results herein are general and they can be extended to many other set-ups.

\begin{figure}[h]
	\begin{center}
		\includegraphics[width=6cm,height=5cm,keepaspectratio]{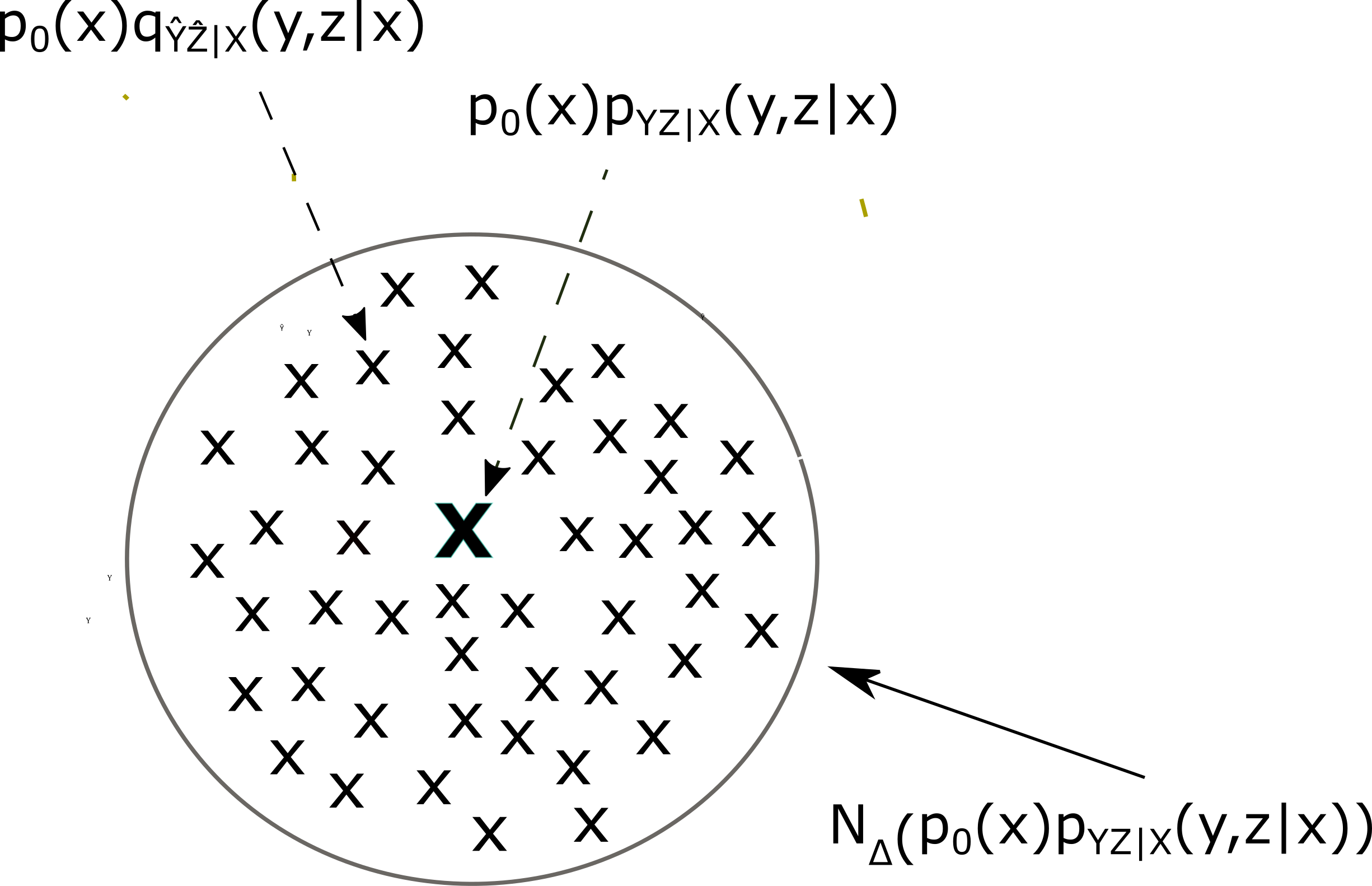}
		\caption{ Achievability part: Every good coordination code designed for achieving perfect empirical coordination according to some distribution  $p_0\left(x\right)q_{\hat{Y},\hat Z|X}\left(y,z|x\right)\in N_{\Delta}\big(p_0\left(x\right)p_{Y,Z|X}\left(y,z|x\right)\big)$ achieves $\Delta$-empirical coordination according to $p_0\left(x\right)p_{Y,Z|X}\left(y,z|x\right)$.}
		\label{fig:fig3}	
	\end{center}
\end{figure}

Our achievability part states that
every good coordination code designed for achieving perfect empirical coordination according to any distribution in $N_{\Delta}\big(p_0\left(x\right)p_{YZ|X}\left(y,z|x\right)\big)$ achieves $\Delta$-empirical coordination according to $p_0\left(x\right)p_{YZ|X}\left(y,z|x\right)$ (see Fig. \ref{fig:fig3}.) 
This means that, if $\left(R_1,R_2,q_{\hat{Y},\hat{Z}|X}\right) \in C_{p_0}^P$ for some $p_0\left(x\right)q_{\hat{Y}\hat Z|X}\left(y,z|x\right) \in N_{\Delta}\big(p_0\left(x\right)p_{YZ|X}\left(y,z|x\right)\big)$, then,
$\left(R_1,R_2,\Delta\right) \in R_{p_0}^I\big(p_{YZ|X}\left(y,z|x\right)\big)$.

\begin{figure}[h]
	\begin{center}
		\includegraphics[width=6cm,height=5cm,keepaspectratio]{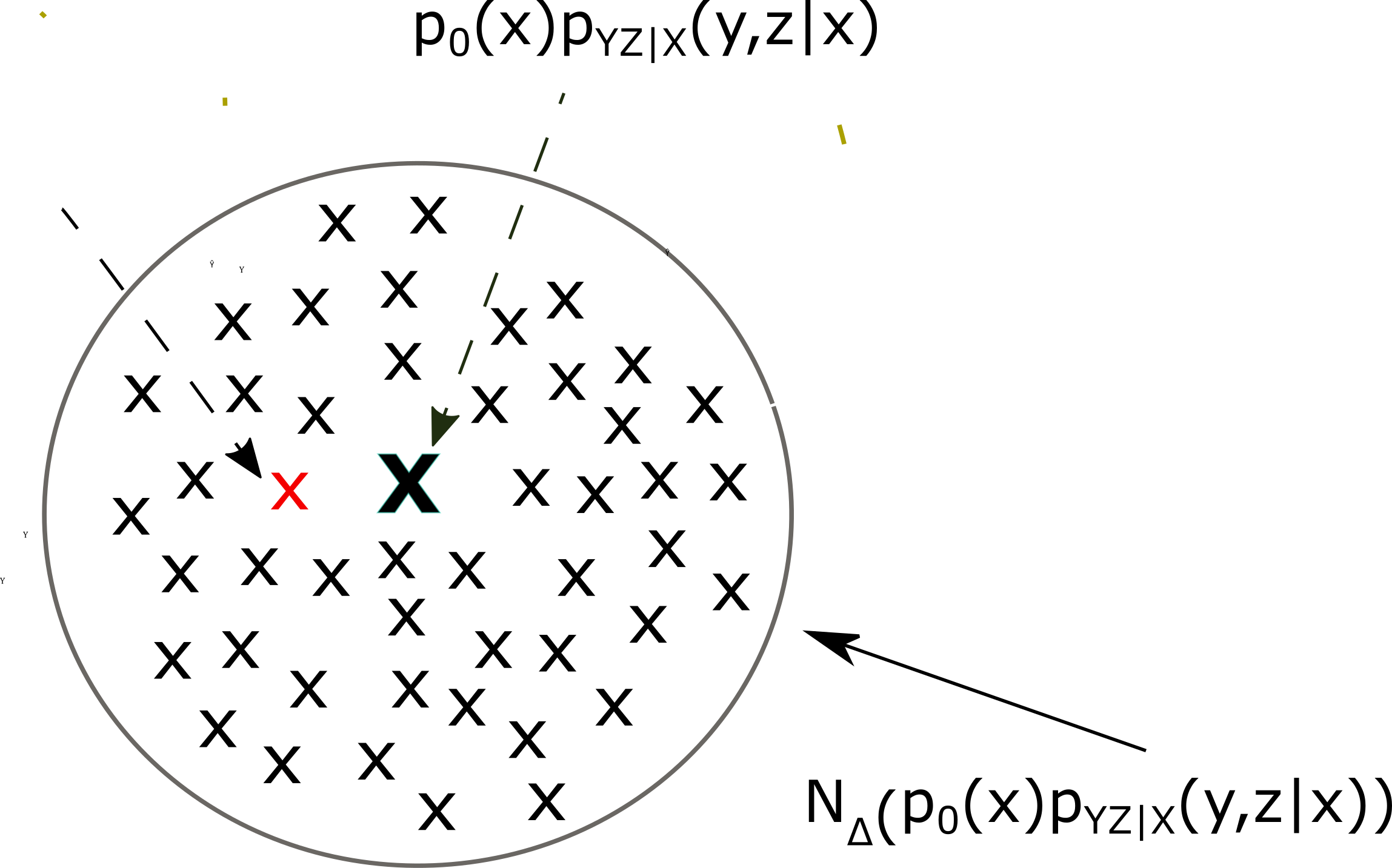}
		\caption{Converse part: For every coordination code that achieves $\Delta$-empirical coordination according to $p_0\left(x\right)p_{Y,Z|X}\left(y,z|x\right)$, there is a coordination code with the same rates which achieves perfect empirical coordination according to some distribution in $N_{\Delta}\big(p_0\left(x\right)p_{Y,Z|X}\left(y,z|x\right)\big)$ (red).}
		\label{fig:fig4}	
	\end{center}
\end{figure}

Our main result makes a stronger statement. According to the converse part,
there is no  more efficient way of satisfying the $\Delta$ coordination-distortion limit than by using a coordination code that produces actions of the same joint type which belongs to $N_{\Delta}\big(p_0\left(x\right)p_{Y,Z|X}\left(y,z|x\right)\big)$. Clearly, a good coordination code, designed for imperfect empirical coordination would produce a variety of different joint types, satisfying on the average the distortion limit.
However, given such a coordination code, repeated uses will produce a longer coordination code with the same rates that achieves perfect empirical coordination according to the expected joint type. This expected joint type can be shown to belong to $N_{\Delta}\big(p_0\left(x\right)p_{Y,Z|X}\left(y,z|x\right)\big)$ and therefore according to the achievability part this new code achieves the distortion limit. Since the two codes have the same rates, we do not lose in rate-efficiency if we substitute the old by the new one (see Fig. \ref{fig:fig4}). This further means that, if $\left(R_1,R_2,\Delta\right) \in R_{p_0}^I\big(p_{Y,Z|X}\left(y,z|x\right)\big)$, then, $\left(R_1,R_2,q_{\hat{Y},\hat{Z}|X}\right) \in C_{p_0}^P$ for some $p_0\left(x\right)q_{\hat{Y},\hat Z|X}\left(y,z|x\right)$ in $ N_{\Delta}\big(p_0\left(x\right)p_{Y,Z|X}\left(y,z|x\right)\big)$.

As a consequence, when we are interested in designing a good coordination code that achieves $\Delta$-empirical coordination according to some distribution, we can restrict our search, without loss of optimality, exactly to the class of coordination codes that achieve perfect empirical coordination according to any distribution in the $\Delta$-neighborhood of this desired distribution. 

The previous discussion is formalized in the next theorem.

\begin{theorem}
	For every $p_{Y,Z|X}\left(y,z|x\right)$,	
	
	\begin{multline*}
	R_{p_0}^I\big(p_{Y,Z|X}\left(y,z|x\right)\big)\\\quad=
	\left.
	\left\{ \,
	\begin{IEEEeqnarraybox}[
	\IEEEeqnarraystrutmode
	\IEEEeqnarraystrutsizeadd{2pt}
	{2pt}][c]{l}
	\left(R_1,R_2,\Delta\right):\\
	\left(R_1,R_2,q_{\hat{Y},\hat{Z}|X}\right)\in C_{p_0}^P\quad\\ \text{for some $\left(\hat Y,\hat Z\right)$ which satisfy}\\ p_0\left(x\right)q_{\hat{Y},\hat Z|X}\left(y,z|x\right)\in N_{\Delta}\big(p_0\left(x\right)p_{Y,Z|X}\left(y,z|x\right)\big)
	\end{IEEEeqnarraybox}\right\}.
	\right.
	%\label{eq:example_left_right2}.
	\end{multline*}
		\label{th:maintheorem}	
\end{theorem}
\begin{remark} Note that in Theorem \ref{th:maintheorem},
$N_{\Delta}\big(p_0\left(x\right)p_{Y,Z|X}\left(y,z|x\right)\big)$ is non-empty since $p_0\left(x\right)p_{Y,Z|X}\left(y,z|x\right)\in N_{\Delta}\big(p_0\left(x\right)p_{Y,Z|X}\left(y,z|x\right)\big)$.
\end{remark}
\begin{IEEEproof}
	%[Proof of the theorem]
	\textbf{Achievability}. %We choose any distribution $q_{\hat Y\hat{Z}|X}\left(y,z|x\right)$ such that $ p_0\left(x\right)q_{\hat{Y}\hat Z|X}\left(y,z|x\right)\in  N_{\Delta}\big(p_0\left(x\right)p_{YZ|X}\left(y,z|x\right)\big)$.
	When a rate-coordination tuple $\left(R_1,R_2,q_{\hat{Y},\hat{Z}|X}\right)$ is in the interior of $C_{p_0}^P$ for some $q_{\hat Y,\hat{Z}|X}\left(y,z|x\right)$ such that $ p_0\left(x\right)q_{\hat{Y},\hat Z|X}\left(y,z|x\right)\in  N_{\Delta}\big(p_0\left(x\right)p_{Y,Z|X}\left(y,z|x\right)\big)$, we are assured (see Definition \ref{def:perfect})
	the existence of a  coordination code such  that for every $\epsilon>0$ and for all large enough $n$ we have
	\begin{equation*}\Pr\left\{\|P_{x^n,y^n,z^n}\left(x,y,z\right)-p_{0}\left(x\right)q_{\hat Y,\hat Z|X}\left(y,z|x\right)\|_{TV}>\epsilon\right\}<\epsilon.
	\end{equation*}
	Applying the triangle inequality on the total variation and because  $ p_0\left(x\right)q_{\hat{Y},\hat Z|X}\left(y,z|x\right)\in  N_{\Delta}\big(p_0\left(x\right)p_{Y,Z|X}\left(y,z|x\right)\big)$, we obtain
	\begin{IEEEeqnarray*}{rCl}\IEEEeqnarraymulticol{3}{l}{\|P_{x^n,y^n,z^n}\left(x,y,z\right)-p_{0}\left(x\right)p_{Y,Z|X}\left(y,z|x\right)\|_{TV}} \\ &\leq&\|P_{x^n,y^n,z^n}\left(x,y,z\right)-p_{0}\left(x\right)q_{\hat Y,\hat Z|X}\left(y,z|x\right)\|_{TV}\\ &\quad\quad +&\|p_{0}\left(x\right)q_{\hat Y,\hat Z|X}\left(y,z|x\right)-p_{0}\left(x\right)p_{Y,Z|X}\left(y,z|x\right)\|_{TV}\\&\leq& \|P_{x^n,y^n,z^n}\left(x,y,z\right)-p_{0}\left(x\right)q_{\hat Y,\hat Z|X}\left(y,z|x\right)\|_{TV}+\Delta\nonumber.	\end{IEEEeqnarray*}
	%\\&\leq\frac{\epsilon}{2}+\Delta
	%\|p_{0}\left(x\right)q\left(y|x\right)-p_{0}\left(x\right)p\left(y|x\right)\|_{TV}
	Thus, for every $\epsilon>0$ and for all large enough $n$, this coordination code achieves
	
	%\begin{multline*}
			%\Pr\left\{\|P_{x^n,y^n,z^n}\left(x,y,z\right)-p_{0}\left(x\right)p_{YZ|X}\left(y,z|x\right)\|_{TV}>\Delta+\epsilon\right\}\\<\epsilon
%	\end{multline*}
\begin{multline*}
\Pr\Big\{\|P_{x^n,y^n,z^n}\left(x,y,z\right)-p_{0}\left(x\right)p_{Y,Z|X}\left(y,z|x\right)\|_{TV}\\ >\Delta+\epsilon\Big\}<\epsilon,
\end{multline*}
%	\begin{IEEEeqnarray*}{rCl}\IEEEeqnarraymulticol{3}{l}{
		%	\Pr\left\{\|P_{x^n,y^n,z^n}\left(x,y,z\right)-p_{0}\left(x\right)p_{YZ|X}\left(y,z|x\right)\|_{TV}>\Delta+\epsilon\right\}}\\&<&\epsilon
	%\end{IEEEeqnarray*}
    which gives 
	\begin{IEEEeqnarray*}{rCl}\IEEEeqnarraymulticol{3}{l}{
			\mathbb{E}\Big\{\|P_{x^n,y^n,z^n}\left(x,y,z\right)-p_{0}\left(x\right)p_{Y,Z|X}\left(y,z|x\right)\|_{TV}\Big\}}\\&\leq&\Pr\Big\{\|P_{x^n,y^n,z^n}\left(x,y,z\right)-p_{0}\left(x\right)p_{Y,Z|X}\left(y,z|x\right)\|_{TV}\\&&\quad\quad\quad\quad\quad\quad\quad\quad\quad\quad\quad\quad\quad>\Delta+\epsilon\Big\}\times TV_{\max}\\ &+&\Pr\Big\{\|P_{x^n,y^n,z^n}\left(x,y,z\right)-p_{0}\left(x\right)p_{Y,Z|X}\left(y,z|x\right)\|_{TV}\\&&\quad\quad\quad\quad\quad\quad\quad\quad\quad\quad\quad\quad\quad\leq\Delta+\epsilon\Big\}\times \left(\Delta+\epsilon\right)\\&\leq& \Delta+\epsilon+\epsilon TV_{\max}.
	\end{IEEEeqnarray*}
	By choosing $\epsilon$ arbitrarily small and $n$ large enough, we conclude that this coordination code achieves also $\Delta$-empirical coordination according to $p_{Y,Z|X}\left(y,z|x\right)$ and so $\left(R_1,R_2,\Delta\right)\in R_{p_0}^I\big(p_{Y,Z|X}\left(y,z|x\right)\big)$.
	
\textbf{Converse}. Suppose that $\left(R_1,R_2,\Delta\right)$ is in the interior of  $R_{p_0}^I\big(p_{Y,Z|X}\left(y,z|x\right)\big)$, i.e., there exists  a coordination code with blocklength $n$ large enough which achieves $\Delta$-empirical coordination according to $p_{Y,Z|X}\left(y,z|x\right)$, at rates $\left(R_1,R_2\right)$ such that
	\begin{IEEEeqnarray*}{c}\mathbb{E}\big\{\|P_{x^n,y^n,z^n}\left(x,y,z\right)-p_{0}\left(x\right)p_{Y,Z|X}\left(y,z|x\right)\|_{TV}\big\}\leq\Delta.
	\end{IEEEeqnarray*}
   This coordination code induces a distribution $\tilde{p}\left(x^n,y^n,z^n\right)=p_{X^n}\left(x^n\right)q_{\hat{ Y}^n,\hat{Z}^n|X^n}\left(y^n,z^n|x^n\right)$ where $p_{X^n}\left(x^n\right)=\prod_{i=1}^{n}{p_0\left(x_i\right)}$.
	%Since, $\mathbf X$ is random, also $\hat{\mathbf Y},\hat{\mathbf Z}$ are random.
	We repeat the use of the scheme over $k$ blocks of length $n$ each and, as a result, we induce a joint distribution on $\left(X^{kn},\hat{Y}^{kn},\hat{Z}^{kn}\right)$ that consists of blocks $\left(X_1^{n},\hat{Y}_1^{n},\hat{Z}_1^{n}\right),\dots,\left(X_{kn-n+1}^{kn},\hat{Y}_{kn-n+1}^{kn},\hat{Z}_{kn-n+1}^{kn}\right)$
	denoted as  $\left(X^{\left(1\right)n},\hat{Y}^{\left(1\right)n},\hat{Z}^{\left(1\right)n}\right),\dots,\left(X^{\left(k\right)n},\hat{Y}^{\left(k\right)n},\hat{Z}^{\left(k\right)n}\right)$. The new coding scheme has rates $R^\prime_i=\frac{\log\big(\left(2^{nR_i}\right)^k\big)}{kn}=R_i$, for $i=1,2$.
	%again, $\left(R_1,R_2\right)$.
	By the law of large numbers, we get
	\begin{IEEEeqnarray*}{c}
		P_{x^{kn},y^{kn},z^{kn}}=\frac{1}{k}\sum_{i=1}^{k}{P_{x^{\left(i\right)n},y^{\left(i\right)n},z^{\left(i\right)n}}}\to \mathbb E\left\{P_{x^n,y^n,z^n}\right\},
	\end{IEEEeqnarray*}
	in probability.
	Point-wise convergence in probability further implies that as $k$ grows
	\begin{IEEEeqnarray*}{c}
		\|P_{x^{kn},y^{kn},z^{kn}}\left(x,y,z\right)-\mathbb E\left\{P_{x^n,y^n,z^n}\right\}\|_{TV}\to 0,
	\end{IEEEeqnarray*}
	in probability.
	However,
	\begin{IEEEeqnarray*}{c}
		\mathbb E\left\{P_{x^n,y^n,z^n}\right\}=p_{\overbar{X},\overbar{\hat{Y}},\overbar{\hat{Z}}}(x,y,z),
	\end{IEEEeqnarray*}
    and thus
    \begin{IEEEeqnarray*}{c}
    \|P_{x^{kn},y^{kn},z^{kn}}\left(x,y,z\right)-p_{\overbar{X},\overbar{\hat{Y}},\overbar{\hat{Z}}}(x,y,z)\|_{TV}\to 0,
    \end{IEEEeqnarray*}
    in probability,
    where 
	\begin{align*}p_{\overbar{X},\overbar{\hat{Y}},\overbar{\hat{Z}}}\left(x,y,z\right)&\triangleq \sum_{k=1}^{n}{\frac{p_{0k}\left(x\right)q_{\hat{Y}_k,\hat Z_k|X_k}\left(y,z|x\right)}{n}} \\&=\sum_{k=1}^{n}{\frac{p_0\left(x\right)q_{\hat{Y}_k\hat Z_k|X_k}\left(y,z|x\right)}{n}}\\&=p_0\left(x\right)\underbrace{\sum_{k=1}^{n}{\frac{q_{\hat{Y}_k,\hat{Z}_k|X_k}\left(y,z|x\right)}{n}}}_{\triangleq p_{\overbar{\hat{Y}},\overbar{\hat{Z}}|\overbar{X}}},
	\end{align*}
	(see  Lemma \ref{lem:mainlem} in the Appendix). %\ref{lem:mainlem}.
	Moreover,
	\begin{align*} \Delta&=\mathbb{E}\big\{\|P_{x^n,y^n,z^n}\left(x,y,z\right)-p_{0}\left(x\right)p_{Y,Z|X}\left(y,z|x\right)\|_{TV}\big\}\\&\stackrel{(a)}{\geq} \|\mathbb{E}\left\{P_{x^n,y^n,z^n}\right\}-p_{0}\left(x\right)p_{Y,Z|X}\left(y,z|x\right)\|_{TV} \\
	&=\|p_{\overbar{X}\overbar{\hat Y}\overbar{\hat Z}}\left(x,y,z\right)-p_{0}\left(x\right)p_{Y,Z|X}\left(y,z|x\right)\|_{TV}\\	&=\|p_{0}\left(x\right)p_{\overbar{\hat{Y}},\overbar{\hat{Z}}|\overbar{X}}\left(y,z|x\right)-p_{0}\left(x\right)p_{Y,Z|X}\left(y,z|x\right)\|_{TV},\end{align*}
	where ($a$) follows from Jensen's inequality since total variation is convex, i.e., for every $\lambda\in \left[0,1\right]$ we have
	\begin{multline*}\lambda\|p_1\left(x,y\right)-q\left(x,y\right)\|_{TV}+\left(1-\lambda\right)\|p_2\left(x,y\right)-q\left(x,y\right)\|_{TV}\\\geq\|\big(\lambda p_1\left(x,y\right)+\left(1-\lambda\right)p_2\left(x,y\right)\big)-q\left(x,y\right)\|_{TV}.\end{multline*}
	%(Lemma \ref{lem:totalv}).
	Thus, we have constructed a sequence of coordination codes with rates $\left(R_1,R_2\right)$ that achieves perfect empirical coordination according to $p_{\overbar{X},\overbar{\hat{Y}},\overbar{\hat{Z}}}=p_{0}\left(x\right)p_{\overbar{\hat{Y}},\overbar{\hat{Z}}|\overbar{X}}\left(y,z|x\right)$, i.e., $\left(R_1,R_2,p_{\overbar{\hat{Y}},\overbar{\hat{Z}}|\overbar{X}}\right)\in C_{p_0}^P $ and additionally
	$p_0\left(x\right)p_{\overbar{\hat{Y}},\overbar{\hat{Z}}|\overbar{X}}\left(y,z|x\right)\in N_\Delta\left(p_0\left(x\right)p_{Y,Z|X}\left(y,z|x\right)\right)$. This completes the proof.
	% $p_0\left(x\right)p_{\overbar{\hat{Y}}\overbar{\hat{Z}}|X}\left(y,z|x\right)\in  N_{\Delta}\big(p_0\left(x\right)p_{YZ|X}\left(y,z|x\right)\big)$.%
	% So, we conclude that $\left(R_1,R_2,p_{\overbar{\hat{Y}}\overbar{\hat{Z}}|X}\left(y,z|x\right)\right)\in C_{p_0}^P$.
	%so it's clear that if we coordinate perfectly the nodes Y,Z to $q_{\hat Y\hat Z|X}\left(y,z|x\right)$, the first term tends to 0  and we achieve our goal.	
	%\begin{multline*}\uplim_{n \to \infty }\mathbb{E}\big\{\|P_{x^n,y^n,z^n}\left(x,y,z\right)-p_{0}\left(x\right)p_{YZ|X}\left(y,z|x\right)\|_{TV}\big\}\\ \leq\lim_{n\to \infty}\mathbb E\Big\{\|P_{x^n,y^n,z^n}\left(x,y,z\right)-p_{0}\left(x\right)q_{\hat Y\hat Z|X}\left(y,z|x\right)\|_{TV}\Big\}+\Delta\nonumber  \end{multline*}
\end{IEEEproof}

\section{Examples}
\begin{figure}[h]
	\begin{center}
		\includegraphics[width=4cm,height=2cm,keepaspectratio]{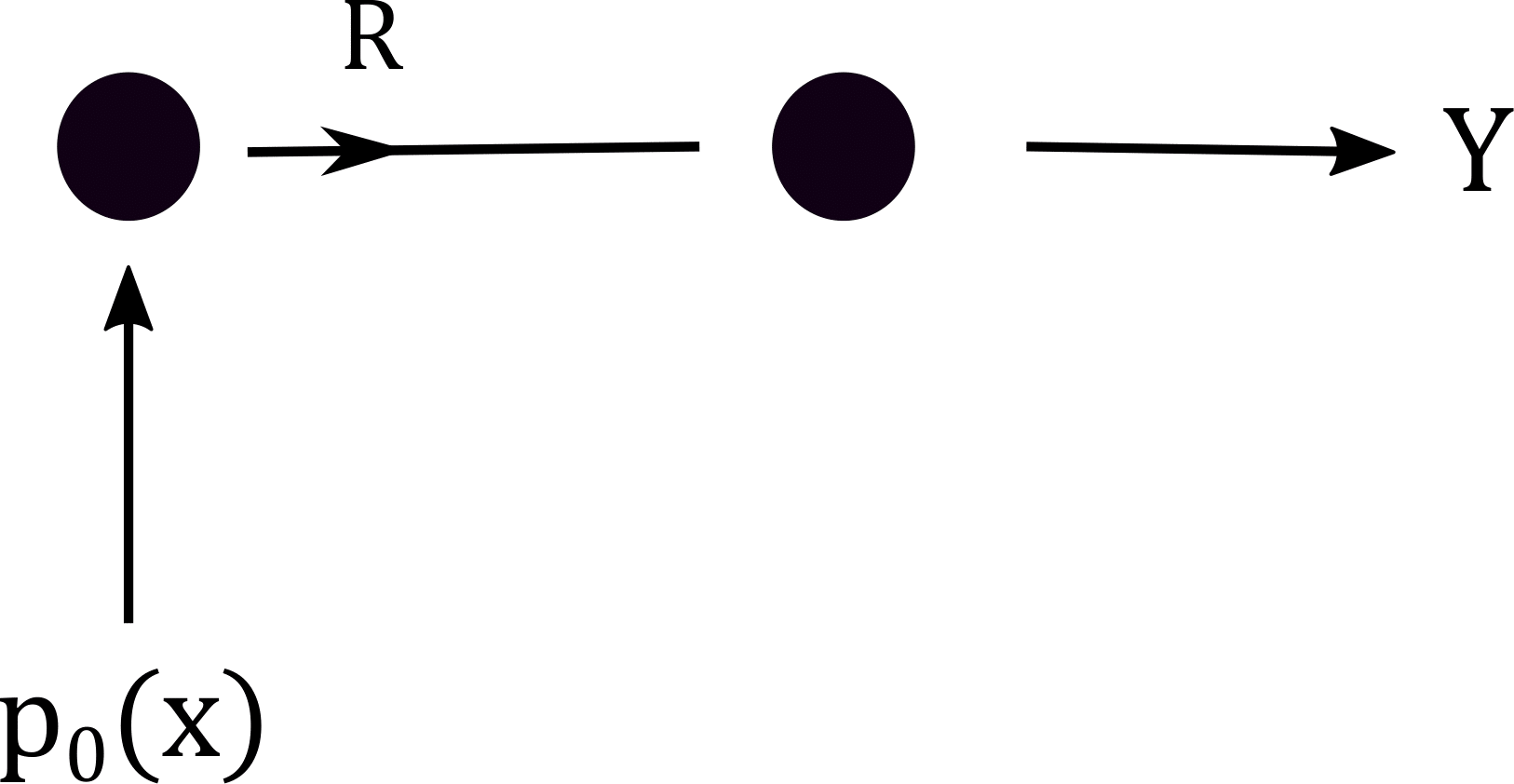}
		\caption{Two node-network.}
		\label{fig:fig5}
	\end{center}
\end{figure}

\begin{figure}[h]
	\begin{center}
		\includegraphics[width=5cm,height=3cm,keepaspectratio]{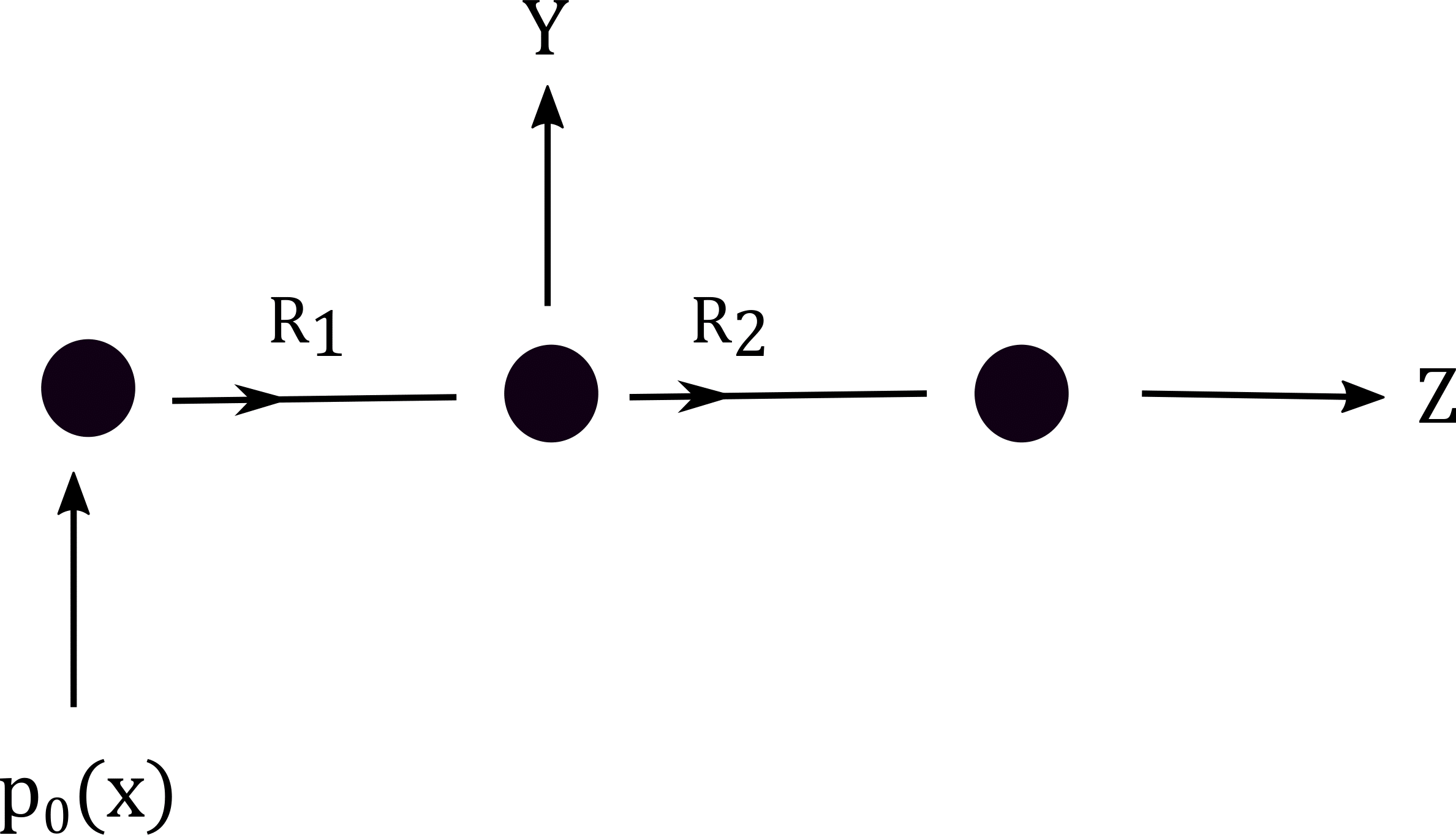}
		\caption{Cascade network.}
		\label{fig:fig6}
	\end{center}
\end{figure}

In this section, we apply Theorem \ref{th:maintheorem} to get the rate-distortion-coordination region in the two node-network illustrated in Fig. \ref{fig:fig5} and in the cascade network illustrated in Fig. \ref{fig:fig6}. 
%The goal is to compare known results on the coordination capacity regions achieved using perfect coordination with the coordination capacity obtained using imperfect empirical coordination.
\begin{lemma}
	Consider the setup of Fig. \ref{fig:fig5}. Then,
	\begin{equation*} C_{p_0}^P=\big\{\big(R,p_{Y|X}\left(y|x\right)\big):R\geq I\left(X;Y\right)\big\}.
	\end{equation*}	
	\label{example1}
\end{lemma}
\begin{IEEEproof}See \cite[Theorem 3]{cuff:2010}.
\end{IEEEproof}
\begin{lemma}
Consider the setup of Fig. \ref{fig:fig5}. Then,
	\begin{multline*}
	R_{p_0}^I\big(p_{Y|X}\left(y|x\right)\big)\\=
	\left.
	\left\{ \,
	\begin{IEEEeqnarraybox}[
	\IEEEeqnarraystrutmode
	\IEEEeqnarraystrutsizeadd{1pt}
	{1pt}][c]{l}
	\left(R,\Delta\right):\\
	R\geq \min_{\substack{q_{\hat{Y}|X}\left(y|x\right):\\ p_0\left(x\right)q_{\hat{Y}|X}\left(y|x\right)\in  N_{\Delta}\big(p_0\left(x\right)p_{Y|X}\left(y|x\right)\big)}} I\left(X;\hat{Y}\right)
	\end{IEEEeqnarraybox}\right\}
	\right.,
	\end{multline*}	
	for every $p_{Y|X}\left(y|x\right)$.
	\label{example2}
\end{lemma}	
\begin{IEEEproof} From Theorem \ref{th:maintheorem}, we obtain
	
	\begin{multline*}
	R_{p_0}^I\big(p_{Y|X}\left(y|x\right)\big)\\=
	\left.
	\left\{ \,
	\begin{IEEEeqnarraybox}[
	\IEEEeqnarraystrutmode
	\IEEEeqnarraystrutsizeadd{1pt}
	{1pt}][c]{l}
	\left(R,\Delta\right):\\
	\left(R,q_{\hat{Y}|X}\right)\in C_{p_0}^P \quad\\ \text{for some $\hat Y$ which satisfy}\\ p_0\left(x\right)q_{\hat{Y}|X}\left(y|x\right)\in N_{\Delta}\big(p_0\left(x\right)p_{Y|X}\left(y|x\right)\big)
	\end{IEEEeqnarraybox}\right\}.
	\right.
	\end{multline*}
	Using Lemma \ref{example1}, the latter can be reformulated as
	\begin{multline*}
	R_{p_0}^I\big(p_{Y|X}\left(y|x\right)\big)\\\quad=
	\left.
	\left\{ \,
	\begin{IEEEeqnarraybox}[
	\IEEEeqnarraystrutmode
	\IEEEeqnarraystrutsizeadd{1pt}
	{1pt}][c]{l}
	\left(R,\Delta\right):\\R\geq I\left(X;\hat{Y}\right)
	\\ \text{for some $\hat Y$ which satisfy}\\ p_0\left(x\right)q_{\hat{Y}|X}\left(y|x\right)\in N_{\Delta}\big(p_0\left(x\right)p_{Y|X}\left(y|x\right)\big)
	\end{IEEEeqnarraybox}\right\}
	\right.,
	%\label{eq:example_left_right2}.
	\end{multline*}
  and by taking into account the non-emptiness and the closeness of $N_{\Delta}\big(p_0\left(x\right)p_{Y|X}\left(y|x\right)\big)$, we obtain the characterization of the lemma.
	%	\begin{multline*}
	%R_{p_0}^I\big(p_{Y|X}\left(y|x\right)\big)\\=
	%\left.
	%\left\{ \,
%	\begin{IEEEeqnarraybox}[
%	\IEEEeqnarraystrutmode
%	\IEEEeqnarraystrutsizeadd{1pt}
	%{1pt}][c]{l}
	%\left(R,\Delta\right):\\
	%R\geq \min_{\substack{q_{\hat{Y}|X}\left(y|x\right): \\ p_0\left(x\right)q_{\hat{Y}|X}\left(y|x\right)\in  N_{\Delta}\big(p_0\left(x\right)p_{Y|X}\left(y|x\right)\big)}} I\left(X;\hat{Y}\right)
%	\end{IEEEeqnarraybox}\right\}.
%	\right.
	%\end{multline*}	
%where we took into account the non-emptiness and the closeness of %$N_{\Delta}\big(p_0\left(x\right)p_{Y|X}\left(y|x\right)\big)$.
\end{IEEEproof}
%\begin{equation*} R_{p_0}^I\big(p_{Y|X}\left(y|x\right)\big)=\Bigg\{\left(R,\Delta\right):R\geq \underbrace{\inf_{q_{\hat{Y}|X}\left(y|x\right):p_0\left(x\right)q_{\hat{Y}|X}\left(y|x\right)\in  N_{\Delta}\big(p_0\left(x\right)p_{Y|X}\left(y|x\right)\big)} I\left(X;\hat{Y}\right)}_{R^f\left(\Delta\right)}\Bigg\}
%\end{equation*}

\begin{lemma}
	Consider the set-up of Fig. \ref{fig:fig6}. Then,
	\begin{equation*} C_{p_0}^P=\left\{\,\begin{IEEEeqnarraybox}[
	\IEEEeqnarraystrutmode
	\IEEEeqnarraystrutsizeadd{1pt}
	{1pt}][c]{l}
	\big(R_1,R_2,p_{Y,Z|X}\left(y,z|x\right)\big):\\
	R_1\geq I\left(X;Y,Z\right),R_2\geq I\left(X;Z\right)
	\end{IEEEeqnarraybox}\right\}.
	%\right.
	\end{equation*}	
	\label{example3}
\end{lemma}
\begin{IEEEproof} See \cite[Theorem 5]{cuff:2010}.
\end{IEEEproof}

\begin{lemma}
	Consider the set-up of Fig. \ref{fig:fig6}. Then,
	\begin{multline*}
	R_{p_0}^I\left(p_{Y,Z|X}\left(y,z|x\right)\right)\\\quad=
	\left.
	\left\{ \,
	\begin{IEEEeqnarraybox}[
	\IEEEeqnarraystrutmode
	\IEEEeqnarraystrutsizeadd{1pt}
	{1pt}][c]{l}
	\big(R_1,R_2,\Delta\big):\\
	R_1\geq I\left(X;\hat Y,\hat{Z}\right),R_2\geq I\left(X;\hat{Z}\right)\\\text{for some $\left(\hat Y,\hat Z\right)$ which satisfy}\\ p_0\left(x\right)q_{\hat{Y},\hat Z|X}\left(y,z|x\right)\in  N_{\Delta}\big(p_0\left(x\right)p_{Y,Z|X}\left(y,z|x\right)\big)
	\end{IEEEeqnarraybox}\right\}
	\right.
	\end{multline*}	
	for every $p_{Y,Z|X}\left(y,z|x\right)$.
	\label{example4}
\end{lemma}
\begin{IEEEproof} From Theorem \ref{th:maintheorem}, we obtain
	\begin{multline*}
	R_{p_0}^I\big(p_{Y,Z|X}\left(y,z|x\right)\big)\\\quad=
	\left.
	\left\{ \,
	\begin{IEEEeqnarraybox}[
	\IEEEeqnarraystrutmode
	\IEEEeqnarraystrutsizeadd{1pt}
	{1pt}][c]{l}
	\left(R_1,R_2,\Delta\right):\\
	\left(R_1,R_2,q_{\hat{Y},\hat{Z}|X}\right)\in C_{p_0}^P\quad\\ \text{for some $\left(\hat Y,\hat Z\right)$ which satisfy}\\ p_0\left(x\right)q_{\hat{Y},\hat Z|X}\left(y,z|x\right)\in N_{\Delta}\big(p_0\left(x\right)p_{Y,Z|X}\left(y,z|x\right)\big)
	\end{IEEEeqnarraybox}\right\}
	\right.,
	%\label{eq:example_left_right2}.
	\end{multline*}	
	and by applying Lemma \ref{example3}, we obtain the characterization of the lemma.\end{IEEEproof}
	%R_{p_0}^I\left(p_{YZ|X}\left(y,z|x\right)\right)\\\quad=
	%\left.
	%\left\{ \,
	%\begin{IEEEeqnarraybox}[
	%\IEEEeqnarraystrutmode
	%\IEEEeqnarraystrutsizeadd{1pt}
	%{1pt}][c]{l}
	%\big(R_1,R_2,\Delta\big):\\
	%R_1\geq I\left(X;\hat Y,\hat{Z}\right),R_2\geq I\left(X;\hat{Z}\right)\\\text{for some $\left(\hat Y,\hat Z\right)$ which satisfy}\\ p_0\left(x\right)q_{\hat{Y}\hat Z|X}\left(y,z|x\right)\in  N_{\Delta}\big(p_0\left(x\right)p_{YZ|X}\left(y,z|x\right)\big)
	%\end{IEEEeqnarraybox}\right\}.
	%\right.
%	\end{multline*}	

%\begin{align*} 
%R_{p_0}^I\left(p_{YZ|X}\left(y,z|x\right)\right)=\Bigg\{\big(R_1,R_2,\Delta\big):\quad & R_1\geq I\left(X;\hat Y,\hat{Z}\right),R_2\geq I\left(X;\hat{Z}\right)\\&\text{for some $\left(\hat Y,\hat Z\right)$ which satisfy}\\& p_0\left(x\right)q_{\hat{Y}\hat Z|X}\left(y,z|x\right)\in  N_{\Delta}\big(p_0\left(x\right)p_{YZ|X}\left(y,z|x\right)\big)\Bigg\}
%.\end{align*}

%\begin{equation*} %C_{p_0}^P=\Big\{\big(R_1,R_2,p_{YZ|X}\left(y,z|x\right)\big):R_1\g%eq I\left(X;Y,Z\right),R_2\geq I\left(X;Z\right)\Big\}. %\label{th:perfect}
%\end{equation*}

%\section{Conclusion}
%We have studied the general problem of imperfect empirical coordination with an average distortion criterion. We have shown that the designing of a good code for this problem is equivalent to the designing of a code which achieves perfect empirical coordination to a carefully chosen distribution. By applying this argument, we have taken the coordination capacity regions of some simple set-ups. However, our argument is general and it can be applied in more general settings.

\section {Appendix}
\begin{lemma}For a given probability distribution $p_{X^n}\left(x^n\right)$,
	\begin{IEEEeqnarray*}{c}\mathbb{E}\big\{P_{x^n}\left(x\right)\big\}=\frac{1}{n}\sum_{k=1}^n\Big(p_{X_k}\left(x\right)\Big).
	\end{IEEEeqnarray*}
	%and any given sequence of length $n$,
	\label{lem:mainlem}	
\end{lemma}
\begin{IEEEproof}
	\begin{IEEEeqnarray*}{rCl}
		\mathbb{E}\big\{P_{x^n}\left(x\right)\big\}&=&\sum_{x^n}p_{X^n}\left(x^n\right)\frac{n_{x^n}\left(x\right)}{n} \\
		&=&\sum_{x^n}\Bigg(p_{X^n}\left(x^n\right)\frac{1}{n}\sum_{k=1}^n\mathbf{1}\left(x_k=x\right)\Bigg)\\
		&=&\sum_{k=1}^n\sum_{x^n}\Bigg(p_{X^n}\left(x^n\right)\frac{1}{n}\mathbf{1}\left(x_k=x\right)\Bigg)\\
		&=&\frac{1}{n}\sum_{k=1}^n\Big(p_{X_k}\left(x\right)\Big).	\end{IEEEeqnarray*}
\end{IEEEproof}

% Bibliography %=================================================================================

\bibliographystyle{IEEEtran}
\bibliography{string,references}

\end{document}